\documentclass[preprintnumbers,amsmath,amssymb,showpacs,twocolumn]{revtex4}
\usepackage{graphicx}% Include figure files
\usepackage{dcolumn}% Align table columns on decimal point
\usepackage{bm}% bold math

\begin{document}

\title{ Quantum secret sharing between multi-party and multi-party  without entanglement}

\author{Feng-Li Yan$^{1,2}$, Ting Gao$^{2,3,4}$}

\affiliation {$^1$ College of Physics, Hebei Normal University, Shijiazhuang 050016, China\\
$^2$ CCAST (World Laboratory), P.O. Box 8730, Beijing 100080, China\\
$^3$ College of Mathematics and Information Science, Hebei Normal University, Shijiazhuang 050016,
China\\
$^4$ Department of Mathematics, Capital Normal University, Beijing 100037, China }

\date{\today}

\begin{abstract}
We propose a quantum secret sharing protocol between multi-party ($m$ members in group 1)  and multi-party ($n$
members in group 2) using a sequence of single photons. These single photons   are used directly to encode
classical information in a quantum secret sharing process. In this protocol, all members in group 1 directly
encode their respective  keys on the states of single photons via  unitary operations, then the last one (the
$m^{th}$ member of group 1) sends $1/n$ of the resulting qubits to each of group 2. Thus the secret message
shared by all members of group 1 is shared by all members of group 2 in such a way that no subset of each group
 is efficient to read the secret message, but  the entire set (not only group 1 but also group 2)
is. We also show that it is unconditionally secure. This protocol is feasible with present-day techniques.
\end{abstract}


\maketitle

\section{Introduction}

Suppose  two groups such as two government departments, where there are $m$  and $n$ members respectively, want
to correspond with each other, but members of each group do not trust each other. What can they do? Classical
cryptography  gives an answer which is known as secret sharing \cite {Schneier}. It can be used, to guarantee
that no single person or part of each department can read out the secret message, but all members of each  group
can. This means that for security to be breached, all people of one group must  act in concert, thereby making
it more difficult for any single person who wants to gain illegal access to the secret information. It can be
implemented as follows: from his original message, every person ( called sender) of group 1 separately creates
$n$ coded messages and sends each of them to each member (called receiver) of group 2. Each of the encrypted
message contains no information about senders' original message, but the combination of all   coded messages
contains the complete message of group 1. However, either a $(m+n+1)$-th party (an "external" eavesdropper) or
the dishonest member of two groups who can gain access to all senders' transmissions can learn the contents of
their (all senders) message in this classical procedure. Fortunately, quantum secret sharing protocols \cite
{HBB, TZG, Gottesman, NQI} can accomplish distributing information securely where multi-photon entanglement is
employed. Recently, many kinds quantum secret sharing with entanglement have been proposed \cite {KKI, CGL, KBB,
BK, XLDP}. Lance {\it et al.} have reported an experimental demonstration of a (2,3) threshold quantum secret
sharing scheme \cite {LSBSL}.
  The combination of quantum key
distribution (QKD) and classical sharing protocol can realize
secret sharing safely.  Quantum secret sharing protocol provides
for secure secret sharing by enabling one to determine whether an
eavesdropper has been active during the secret sharing procedure.
But it is not easy to implement such multi-party secret sharing
tasks \cite {HBB, KKI}, since the efficiency of preparing even
tripartite or four-partite entangled states is very low \cite
{BPDWZ, PDGWZ}, at the same time  the efficiency of the existing
quantum secret sharing protocols using quantum entanglement can
only approach $50\%$.

More recently, a  protocol for quantum secret sharing without entanglement has been proposed by Guo and Guo
\cite {GG03}. They present an idea to directly encode the qubit of quantum key distribution and accomplish one
splitting a message into many parts to achieve multi-party secret sharing only by product states. The
theoretical efficiency is doubled to approach $100\%$. Br\'{a}dler and Du\v{s}ek have given two protocols for
secret-information splitting among many participants \cite{BD}.

In this paper, we propose a quantum secret sharing scheme
employing single qubits to achieve the aim mentioned above --- the
secret sharing between multi-party ($m$ parties of group 1) and
multi-party ($n$ parties of group 2). That is, instead of giving
his information to any one individual of group 1, each sender to
split his information in such a way that no part members  of group
1 or group 2 have any knowledge of the combination of all senders
(group 1), but all members of each group can jointly determine the
combination of all senders (group 1). The security of our scheme
is based on the quantum no-cloning theory just as the BB84 quantum
key distribution. Comparing with the efficiency $50\%$ limiting
for the existing quantum secret sharing protocols with quantum
entanglement, the present scheme can also be $100\%$ efficient in
principle.

\section{ quantum key sharing between multi-party and multi-party}

Suppose there are $m$ ($m\geq 2$) and $n$ ($n\geq 2$) members in
government department1 and department2, respectively, and Alice1,
Alice2, $\cdots$, Alice$m$, and Bob1, Bob2, $\cdots$, Bob$n$ are
their respective all members. $m$ parties of department1  want
quantum key sharing with $n$ parties of department2 such that
neither one nor part of each department knows the key, but only by
all members' working together can each department determine what
the string (key) is. In this case it is the quantum information
that has been split into $n$ pieces,  no one  of which separately
contains the original information, but whose combination does.

Alice1 begins with $A_1$ and $B_1$, two strings each of $nN$
random classical bits. She then encodes these strings as a block
of $nN$ qubits,
\begin{eqnarray}\label{Alice1}
   |\Psi^1\rangle & = & \otimes_{k=1}^{nN}|\psi_{a^1_kb^1_k}\rangle\nonumber\\
 & = & \otimes_{j=0}^{N-1}|\psi_{a^1_{nj+1}b^1_{nj+1}}\rangle|\psi_{a^1_{nj+2}b^1_{nj+2}}\rangle\cdots
 |\psi_{a^1_{nj+n}b^1_{nj+n}}\rangle,
\end{eqnarray}
where $a^1_k$ is the $k^{th}$ bit of $A_1$ (and similar for $B_1$) and each qubit is one of the four states
\begin{eqnarray}\label{singlequbits}
% \nonumber to remove numbering (before each equation)
  |\psi_{00}\rangle &=& |0\rangle,\label{singlequbit1} \\
  |\psi_{10}\rangle &=& |1\rangle,\\
  |\psi_{01}\rangle &=& |+\rangle=\frac{|0\rangle+|1\rangle}{\sqrt{2}},\\
   |\psi_{11}\rangle &=& |-\rangle=\frac{|0\rangle-|1\rangle}{\sqrt{2}}.\label{singlequbit4}
\end{eqnarray}
The effect of this procedure is to encode $A_1$ in the basis $Z=\{|0\rangle, |1\rangle\}$ or $X=\{|+\rangle,
|-\rangle\}$, as determined by $B_1$. Note that the four states are not all mutually orthogonal, therefore no
measurement can distinguish between  all of them with certainty. Alice1 then sends $|\Psi^1\rangle$ to Alice2
over their public quantum communication channel.

Depending on a string $A_2$ of $nN$ random classical bits which she generates, Alice2 subsequently applies a
unitary transformation $\sigma_0=I=|0\rangle\langle0|+|1\rangle\langle1|$ (if the $k^{th}$ bit $a^2_k$  of $A_2$
is $0$), or $\sigma_1=i\sigma_y=|0\rangle\langle1|-|1\rangle\langle0|$ (if $a^2_k=1$) on each
$|\psi_{a^1_kb^1_k}\rangle$ of the $nN$ qubits she receives from Alice1 such that $|\psi_{a^1_kb^1_k}\rangle$ is
changed into $|\psi^0_{a^2_kb^1_k}\rangle$, and obtains $nN$-qubit product state
$|\Psi^{20}\rangle=\otimes_{k=1}^{nN}|\psi^0_{a^2_kb^1_k}\rangle$. After that, she performs a unitary operator
$I$ (if $b^2_k=0$) or
$H=\frac{1}{\sqrt{2}}(|0\rangle+|1\rangle)\langle0|+\frac{1}{\sqrt{2}}(|0\rangle-|1\rangle)\langle1|$ (if
$b^2_k=1$) on each qubit state $|\psi^0_{a^2_kb^1_k}\rangle$ according to her another random classical bits
string $B_2$, and makes $|\psi^0_{a^2_kb^1_k}\rangle$ to be turned into $|\psi_{a^2_kb^2_k}\rangle$. Alice2
sends Alice3 $|\Psi^2\rangle=\otimes_{k=1}^{nN}|\psi_{a^2_kb^2_k}\rangle$. Similar to Alice2, Alice3 applies
quantum operations on each qubit and sends the resulting $nN$ qubits to Alice4. This procedure goes on until
Alice$m$.

Similarly, Alice$m$ first creates two strings $A_m$ and $B_m$ of $nN$ random classical bits. Then she   makes a
unitary operation $\sigma_0$ (if $a^m_k=0$) or $\sigma_1$ (if $a^m_k=1$) on each qubit state
$|\psi_{a^{m-1}_kb^{m-1}_k}\rangle$. It follows that $|\psi_{a^{m-1}_kb^{m-1}_k}\rangle$ is changed into
$|\psi^0_{a^m_kb^{m-1}_k}\rangle$. After that she applies operator $I$ (if $b^m_k=0$) or $H$ (if $b^m_k=1$) on
the resulting qubit state $|\psi^0_{a^m_kb^{m-1}_k}\rangle$ such that $|\psi^0_{a^m_kb^{m-1}_k}\rangle$ is
turned into $|\psi_{a^m_kb^m_k}\rangle$. Alice$m$ sends  $N$-qubit product states
$|\Psi_1^m\rangle=\otimes_{j=0}^{N-1}|\psi_{a^m_{nj+1}b^m_{nj+1}}\rangle$,
$|\Psi_2^m\rangle=\otimes_{j=0}^{N-1}|\psi_{a^m_{nj+2}b^m_{nj+2}}\rangle$, $\cdots$,
 $|\Psi_n^m\rangle=\otimes_{j=0}^{N-1}|\psi_{a^m_{nj+n}b^m_{nj+n}}\rangle$ of the resulting $nN$-qubit state
$|\Psi^m\rangle=\otimes_{k=1}^{nN}|\psi_{a^m_kb^m_k}\rangle$ to
Bob1, Bob2, $\cdots$, Bob$n$, respectively.

When all Bob1, Bob2, $\cdots$, and Bob$n$ have announced  the
receiving of their strings of $N$ qubits, Alice1, Alice2,
$\cdots$, and Alice$m$ publicly announce the strings $B_1$, $B_2$,
$\cdots$, and $B_m$ one after another, respectively. Note that
$B_1$, $B_2$, $\cdots$, and $B_m$ reveal nothing about $A_i$
($i=1, 2, \cdots, m$), but it is important that all Alice1,
Alice2, $\cdots$, and Alice$m$ not publish their respective $B_1$,
$B_2$, $\cdots$, and $B_m$ until after all Bob1, Bob2, $\cdots$,
and Bob$n$  announce  the reception of the $N$ qubits Alice$m$
sends to them.

Bob1, Bob2, $\cdots$, and Bob$n$ then measure each qubit of their
respective strings in the basis $X$ or $Z$ according to the XOR
result of corresponding bits  of strings $B_1$, $B_2$, $\cdots$,
$B_m$. Since the unitary transformation $\sigma_1=i\sigma_y$ flips
the states in both measuring bases such that
$\sigma_1|0\rangle=-|1\rangle$, $\sigma_1|1\rangle=|0\rangle$,
$\sigma_1|+\rangle=|-\rangle$ and $\sigma_1|-\rangle=-|+\rangle$,
i.e. $I, i\sigma_y$ leave bases $X$ and $Z$ unchanged, but $H$
turns $|0\rangle$, $|1\rangle$, $|+\rangle$ and $|-\rangle$ into
$|+\rangle$, $|-\rangle$, $|0\rangle$ and $|1\rangle$,
respectively, i.e. $ H$ changes bases $X$ and $Z$, so if
$\oplus_{i=2}^m b^i_k=b^2_k\oplus b^3_k\oplus\cdots\oplus
b^m_k=0$, then $|\psi_{a^m_kb^m_k}\rangle$ should be measured in
the same basis with $|\psi_{a^1_kb^1_k}\rangle$; if
$\oplus_{i=2}^m b^i_k=1$, $|\psi_{a^m_kb^m_k}\rangle$ should be
measured in the basis different from $|\psi_{a^1_kb^1_k}\rangle$,
where the symbol $\oplus$ is the addition modulo 2. Therefore, if
$\oplus_{i=2}^m b^i_k=b^1_k$, $|\psi_{a^m_kb^m_k}\rangle$ is
measured in the $Z$ basis, otherwise in the basis $X$. That is, if
$\oplus_{i=1}^m b^i_{nj+l}=0$, then Bob$l$ measures
$|\psi_{a^m_{nj+l}b^m_{nj+l}}\rangle$ in the basis $Z$, otherwise,
he measures in the basis $X$. Moreover, after measurements, Bob$l$
can extract out all Alices's encoding information  $\oplus^m_{i=1}
a^i_{nj+l}$, $j=0,1,2, \cdots, N-1$, for $l=1, 2, \cdots, n$.

 Now all Alices
and Bobs perform some tests to determine how much  noise or eavesdropping happened during their communication.
Alice1, Alice2, $\cdots$, and Alice$m$ select some bits $nj_r+l$ (of their $nN$ bits) at random, and publicly
announce the selection. Here $j_r\in\{j_1, j_2, \ldots, j_{r_0}\}\subset\{j_1, j_2, \ldots, j_{r_0}, j_{r_0+1},
\ldots, j_N\}=\{0,1,2,\ldots, N-1\}$, and $l=1,2,\ldots,n$.  All Bobs and all Alices then publish and compare
the values of these checked bits. If they find too few the XOR results $\oplus_{i=1}^ma^i_{nj_r+l}$ of the
corresponding bits $a^i_{nj_r+l}$ of these checked bits of all Alices and the values of Bob$l$'s checked bits
$|\psi_{a^m_{nj_r+l}b^m_{nj_r+l}}\rangle$ agree, then they abort and re-try the protocol from the start. The XOR
results $\oplus^n_{l=1}(\oplus^m_{i=1}a_{nj_s+l}^i)$  of Bob$l$'s corresponding bits $\oplus^m_{i=1}
a_{nj_s+l}^i$ of the rest unchecked bits $nj_s+l$ of $\{\oplus^m_{i=1} a_{nj+1}^i\}^{N-1}_{j=0}$,
  $\{\oplus^m_{i=1} a_{nj+2}^i\}^{N-1}_{j=0}$, $\cdots$, $\{\oplus^m_{i=1} a_{nj+n}^i\}^{N-1}_{j=0}$
  (or $\otimes_{j=0}^{N-1}|\psi_{a^m_{nj+1}b^m_{nj+1}}\rangle$,
$\otimes_{j=0}^{N-1}|\psi_{a^m_{nj+2}b^m_{nj+2}}\rangle$,
$\cdots$,
 $\otimes_{j=0}^{N-1}|\psi_{a^m_{nj+n}b^m_{nj+n}}\rangle$) can be used as raw keys
for secret sharing between all Alices and all Bobs, where
$j_s=j_{r_0+1}, j_{r_0+2}, \ldots, j_N$.

This protocol is summarized as follows:

M1. Alice1 chooses two  random $nN$-bit strings $A_1$ and $B_1$.
She encodes each data bit of $A_1$ as $\{|0\rangle, |1\rangle\}$
if the corresponding bit of $B_1$ is 0 or $\{|+\rangle,
|-\rangle\}$ if $B_1$ is 1. Explicitly,  she encodes each data bit
0 ( 1 ) of $A_1$ as  $|0\rangle$ ( $|1\rangle$ ) if the
corresponding bit of $B_1$ is 0 or $|+\rangle$ ( $|-\rangle$ ) if
the corresponding bit of $B_1$ is 1, i.e. she encodes each bit
$a_k^1$ of $A_1$ as $|\psi_{a^1_kb^1_k}\rangle$ of
Eqs.(\ref{singlequbit1})-(\ref{singlequbit4}), where $b^1_k$ is
the corresponding bit of $B_1$. Then she sends the resulting
$nN$-qubit state
$|\Psi^1\rangle=\otimes_{k=1}^{nN}|\psi_{a^1_kb^1_k}\rangle$ to
Alice2.

M2.  Alice2 creates two random $nN$-bit strings $A_2$ and $B_2$.
She applies $\sigma_0$ or $\sigma_1$ to each qubit
$|\psi_{a^1_kb^1_k}\rangle$ of $nN$-qubit state $|\Psi^1\rangle$
according to the corresponding bit of $A_2$ being 0 or 1, then she
applies $I$ or $H$ to each qubit of the resulting $nN$-qubit state
depending on the corresponding bit of $B_2$ being 0 or 1. After
this, she sends Alice3 the resulting $nN$-qubit state
$|\Psi^2\rangle$.

M3. Alice$i$ does likewise, $i=3, 4, \cdots, m-1$. Depending on the corresponding bit $a_k^m$ of a random
$nN$-bit string $A_m$, which she generates on her own,  Alice$m$ performs $\sigma_0$ (if $a_k^m=0$)  or
$\sigma_1$ (if $a_k^m=1$) on each qubit of $|\Psi^{m-1}\rangle$. According to a random bit string $B_m$ which
she generates, she subsequently applies $I$ (If the corresponding bit $b_k^m$ of  $B_m$ is 0) or $H$ (if
$b_k^m=1$) on each qubit of the resulting $nN$-qubit state $|\Psi^{m0}\rangle$, which results in $nN$-qubit
state $|\Psi^m\rangle=\otimes_{k=1}^{nN}|\psi_{a^m_kb^m_k}\rangle$. After it, she sends $N$-qubit state
$\otimes_{j=0}^{N-1}|\psi_{a^m_{nj+l}b^m_{nj+l}}\rangle$ to Bob$l$, $1\leq l\leq n$.

M4. Bob1, Bob2, $\cdots$, Bob$n$ receive $N$ qubits,  and announce this fact, respectively.

M5. Alice1, Alice2, $\cdots$, and Alice$m$ publicly announce the
strings $B_1$, $B_2$, $\cdots$, and $B_m$, respectively.

M6. Bob1, Bob2, $\cdots$, and Bob$n$  measure each qubit of their respective strings in the basis $Z$ or $X$
according to the XOR results of corresponding bits  of strings $B_1$, $B_2$, $\cdots$, $B_m$. That is, Bob$l$
measures $|\psi_{a^m_{nj+l}b^m_{nj+l}}\rangle$ in the basis $Z$ (if $\oplus_{i=1}^m b_{nj+l}^i=0$) or in the
basis $X$ (if $\oplus_{i=1}^m b_{nj+l}^i=1$), $j=0, 1, \cdots, N-1$,  $l=1, 2, \cdots, n$.

M7. All Alices select randomly a subset  that will serve as a
check on Eve's interference, and tell all Bobs the bits they
choose.  In the check procedure, all Alices and Bobs are required
to broadcast the values of their checked bits, and compare the XOR
results of the corresponding bits of checked bits of $A_1$, $A_2$,
$\cdots$, $A_m$ and the values of the corresponding bits of Bob1,
Bob2, $\cdots$, and Bob$n$. If more than an acceptable number
disagree, they abort this round of operation and restart from
first step.

M8. The XOR results $\oplus^n_{l=1}(\oplus^m_{i=1}a_{nj_s+l}^i)$
of Bob$l$'s corresponding bits $\oplus^m_{i=1} a_{nj_s+l}^i$ of
the remaining bits $nj_s+l$ of $\{\oplus^m_{i=1}
a_{nj+1}^i\}^{N-1}_{j=0}$,
  $\{\oplus^m_{i=1} a_{nj+2}^i\}^{N-1}_{j=0}$, $\cdots$, $\{\oplus^m_{i=1} a_{nj+n}^i\}^{N-1}_{j=0}$
  (or $\otimes_{j=0}^{N-1}|\psi_{a^m_{nj+1}b^m_{nj+1}}\rangle$,
$\otimes_{j=0}^{N-1}|\psi_{a^m_{nj+2}b^m_{nj+2}}\rangle$,
$\cdots$,
 $\otimes_{j=0}^{N-1}|\psi_{a^m_{nj+n}b^m_{nj+n}}\rangle$) can be used as  key bits for secret sharing
  between all Alices and all Bobs,
 where $j_s=j_{r_0+1}, j_{r_0+2}, \ldots, j_N$.

For example, $m=2$ and $n=3$.  Suppose $A_1=\{1, 0,
0,1,0,1,0,1,1,0,0,0,1,1,1,0,1,0\}$ and
$B_1=\{0,1,0,1,1,0,1,1,0,0,1,0,1,0,1,0,0,1\}$ are two random
18-bit strings of Alice1. Depending on $B_1$, then she encodes
$A_1$ as
$|\Psi^1\rangle=|1\rangle|+\rangle|0\rangle|-\rangle|+\rangle|1\rangle|+\rangle|-\rangle|1\rangle|0\rangle|+\rangle
|0\rangle|-\rangle|1\rangle|-\rangle|0\rangle|1\rangle|+\rangle$.
If Alice2's two strings of random bits are
$A_2=\{1,1,1,0,0,1,1,1,0,0,0,1,0,1,1,0,0,1\}$ and
$B_2=\{1,0,0,1,1,0,0,0,1,1,1,1,0,0,0,1,0,1\}$, she applies
$i\sigma_y$ to the $1^{th}$, $2^{th}$, $3^{th}$, $6^{th}$,
$7^{th}$, $8^{th}$, ${12}^{th}$, ${14}^{th}$, ${15}^{th}$,
${18}^{th}$ qubits of $|\Psi^1\rangle$,  getting
$|\Psi^{20}\rangle=|0\rangle|-\rangle|1\rangle|-\rangle|+\rangle|0\rangle|-\rangle|+\rangle|1\rangle|0\rangle|+\rangle
|1\rangle|-\rangle|0\rangle|+\rangle|0\rangle|1\rangle|-\rangle$,
then she performs $H$ on $1^{th}$, $4^{th}$, $5^{th}$, $9^{th}$,
${10}^{th}$, ${11}^{th}$, ${12}^{th}$, ${16}^{th}$,  ${18}^{th}$
qubits of $|\Psi^{20}\rangle$, obtaining
$|\Psi^2\rangle=\otimes^{18}_{k=1}|\psi_{a^2_kb^2_k}\rangle
=|+\rangle|-\rangle|1\rangle|1\rangle|0\rangle|0\rangle|-\rangle|+\rangle|-\rangle|+\rangle|0\rangle
|-\rangle|-\rangle|0\rangle|+\rangle|+\rangle|1\rangle|1\rangle$.
After that, she sends the 6-qubit states
$|\Psi^2_1\rangle=\otimes^5_{j=0}|\psi_{a^2_{3j+1}b^2_{3j+1}}\rangle=|+\rangle|1\rangle|-\rangle|+\rangle|-\rangle|+\rangle$,
$|\Psi^2_2\rangle=\otimes^5_{j=0}|\psi_{a^2_{3j+2}b^2_{3j+2}}\rangle=|-\rangle|0\rangle|+\rangle|0\rangle|0\rangle|1\rangle$,
and
$|\Psi^2_3\rangle=\otimes^5_{j=0}|\psi_{a^2_{3j+3}b^2_{3j+3}}\rangle=|1\rangle|0\rangle|-\rangle|-\rangle|+\rangle|1\rangle$
to Bob1, Bob2 and Bob3, respectively. When each of Bob1, Bob2 and
Bob3 has received $6$-qubit state and announced the fact, Alice1
and Alice2 publicly inform all Bobs their respective  strings
$B_1$ and $B_2$. Then Bob$l$ measures his qubit state
$|\psi_{a^2_{3j+l}b^2_{3j+l}}\rangle$ in the basis $Z$ if
$b^1_{3j+l}\oplus b^2_{3j+l}=0$ or in basis $X$ if
$b^1_{3j+l}\oplus b^2_{3j+l}=1$, for $j=0, 1, \cdots, 5$,  $l=1,
2, 3$. From this, Bob1, Bob2 and Bob3 derive Alice1 and Alice2's
encoding information  $\{0,1,1,0,1,0\}$, $\{1,0,0,0,0,1\}$ and
$\{1,0,1,1,0,1\}$ of their respective 6-qubit states if no Eve's
eavesdropping exists. If Alice1 and Alice2 choose the $1^{th}$,
$2^{th}$, $3^{th}$, $13^{th}$, $14^{th}$, $15^{th}$ bits as the
check bits, then the XOR results $1\oplus 0\oplus 0$, $1\oplus
0\oplus 1$, $0\oplus 0\oplus 1$, $0\oplus 1\oplus 1$ (or 1, 0, 1,
0) of the corresponding bits of Bob1, Bob2 and Bob3's remaining
bits $\{1,1,0,0\}$, $\{0,0,0,1\}$ and $\{0,1,1,1\}$ are used as
raw keys for secret sharing between two Alices and three Bobs.

Note that $B_1$, $B_2$, $\cdots$, and $B_m$ reveal nothing about $A_i$ ($i=1, 2, \cdots, m$), but it is
important that all Alice1, Alice2, $\cdots$, and Alice$m$ not publish their respective $B_1$, $B_2$, $\cdots$,
and $B_m$ until after all Bob1, Bob2, $\cdots$, and Bob$n$  announce  the reception of the $N$ qubits Alice$m$
sends to them. If all Alices broadcast their respective $B_1$, $B_2$, $\cdots$, and $B_m$ before all Bobs
announce  the reception of the $N$ qubits Alice$m$ sends to them, then either a $(m+n+1)$-th party (an
"external" eavesdropper) or the dishonest member of two groups intercepts  $nN$ qubits state
$|\Psi^m\rangle=\otimes_{k=1}^{nN}|\psi_{a^m_kb^m_k}\rangle$  can learn the contents of their (all senders)
message in this procedure by measuring each qubit in the $Z$ basis (if $\oplus_{i=1}^m b_{nj+l}^i=0$) or in the
$X$ basis (if $\oplus_{i=1}^m b_{nj+l}^i=1$).

 It is necessary for Alice$i$ ($2\leq i\leq m$) applying unitary operation $H$ randomly
on some qubits.  Each sender Alice$i$ encoding string $B_i$ on the
sequence of  states of qubits is to achieve the aim   such that no
one or part of Alice1, $\cdots$ , Alice$m$ can extract some
information of others.
 Case I: Alice$2$ does not encode a random string of $I$ and $H$ on the sequence of single
photons, Alice1 can enforce the intercept-resend strategy  to
extract Alice$2$'s whole information.
  Alice1 can intercept all the single photons and measure them, then resend them.  As the sequence of single
  photons  is prepared by Alice1, Alice1 knows the measuring-basis, and the original state of each photon.
   She uses the same  measuring-basis  when she prepared the photon to measure the photon, and read out
 Alice2's complete secret messages directly. Case II: Alice$i_0$ ($3\leq i_0\leq m$) is the first one who does not encode
 a random string  of $I$ and $H$ on the sequence of single photons, then one of Alice1, Alice2, $\ldots$ ,
  Alice$(i_0-1)$ can
 also enforce the intercept-resend strategy  to extract Alice$i_0$'s whole information by their cooperation.
 Without loss of generality, suppose that
 Alice2 intercepts all the particles that Alice$i_0$ sends. Alice2 can obtain Alice$i_0$'s secret message
  if Alice1, Alice3,
  $\ldots$ , Alice$(i_0-1)$ inform her their respective strings $B_1$,
 $B_3$, $\ldots$ , $B_{i_0-1}$ and $A_1$,
 $A_3$, $\ldots$ , $A_{i_0-1}$.

 This secret sharing protocol between $m$ parties and $n$ parties is almost $100\%$ efficient as all the keys
can be used in the ideal case of no eavesdropping,  while the
quantum secret sharing protocols with entanglement states \cite
{HBB} can be at most $50\%$ efficient in principle. In this
protocol, quantum memory is required to store the qubits which has
been shown available in the present experiment technique \cite
{GG02}. However, if no quantum memory is employed,  all Bobs
measure their qubits before  Alice$i$'s ($1 \leq i\leq m$)
announcement of basis, the efficiency of the present protocol
falls to $50\%$.

Two groups can also realize secret sharing by Alice1 preparing a
sequence of $nN$ polarized single photons such that the $n$-qubit
product state of each $n$ photons is in the basis  $Z$ or $X$ as
determined by $N$-bit string $B_1$, instead that  in the above
protocol. For instance,  (A) Alice$i$ ($1\leq i\leq m$) creates a
random $nN$-bit string $A_i$ and a random $N$-bit string $B_i$,
and Alice1
 encodes her two strings as a block of $nN$  qubits state $|\Phi^1\rangle
=\otimes_{j=1}^N|\phi_{a^1_{n(j-1)+1}b^1_j}\rangle|\phi_{a^1_{n(j-1)+2}b^1_j}\rangle\cdots|\phi_{a^1_{n(j-1)+n}b^1_j}\rangle$,
where each qubit state $|\phi_{a^1_{n(j-1)+l}b^1_j}\rangle$ is one
of $|\phi_{00}\rangle=|0\rangle$, $|\phi_{10}\rangle=|1\rangle$,
 $|\phi_{01}\rangle=|+\rangle$ and $|\phi_{11}\rangle=|-\rangle$. Then Alice1 sends $|\Phi^1\rangle$ to Alice2.
 Alice$i$ ($2\leq i\leq m$) applies $\sigma_0$ or $\sigma_1$ to each qubit state $|\phi_{a^{i-1}_{n(j-1)+l}b^{i-1}_j}\rangle$
 ($1\leq l\leq n$)
 according to the corresponding bit $a^i_{n(j-1)+l}$  of $A_2$ being 0 or 1, then she applies $I$ (if $b^i_j=0$) or $H$
  (if $b^i_j=1$) to each resulting qubit state $|\phi^0_{a^i_{n(j-1)+l}b^i_j}\rangle$. Alice$m$ sends $N$ qubits
  $\otimes_{j=1}^N|\phi_{a^m_{n{(j-1)}+l}b^m_j}\rangle$ of the resulting $nN$ qubits state $|\Phi^m\rangle
=\otimes_{j=1}^N|\phi_{a^m_{n(j-1)+1}b^m_j}\rangle|\phi_{a^m_{n(j-1)+2}b^m_j}\rangle\cdots|\phi_{a^m_{n(j-1)+n}b^m_j}\rangle$
to Bob$l$, $1\leq l\leq n$. After  all Bobs receive their respective $N$ qubits, Alice$i$ announces $B_i$, then
Bob$l$ measures  each of his qubit states $|\phi_{a^m_{n(j-1)+l}b^m_j}\rangle$ in the basis $Z$ if
$\oplus_{i=1}^mb^i_j=0$ or $X$ if $\oplus_{i=1}^mb^i_j=1$, and deduces its value $\oplus_{i=1}^ma^i_{n(j-1)+l}$
if there is no Eve's eavesdropping. A subset of $\{\oplus^n_{l=1}(\oplus^m_{i=1}a^i_{n(j-1)+l})\}^N_{j=1}$ will
serve as a check, passing the test, the unchecked bits of
$\{\oplus^n_{l=1}(\oplus^m_{i=1}a^i_{n(j-1)+l})\}^N_{j=1}$ will take as the raw keys for  secret sharing between
two groups. (B)
 Alice$i$ chooses two random $N$-bit strings $A_i$ and $B_i$, and Alice1 prepares a block of $nN$ qubits state $|\Psi^1\rangle=
 \otimes_{j=1}^N|\psi_{a^1_{j1}b^1_j}\rangle|\psi_{a^1_{j2}b^1_j}\rangle\cdots|\psi_{a^1_{jn}b^1_j}\rangle$,
 where $a^1_{jl}$ is 0 or 1 and $\oplus_{l=1}^na^1_{jl}=a^1_j$.
  Alice$i$ applies unitary
 operation $\sigma_0$  or $\sigma_1$ to each qubit state $|\psi_{a_{jl}^{i-1}b_j^{i-1}}\rangle$ depending on
 the $j$-th bit $a^i_j$ of $A_i$ being 0 or 1,  following it, $I$ or $H$ according to $B_i$, to each
 particle. Bob$l$ measures each of his particles $|\psi_{a^m_{jl}b^m_j}\rangle$ in the basis $Z$ (if $\oplus_{i=1}^mb^i_j=0$)
  or  $X$ (if $\oplus_{i=1}^mb^i_j=1$). All Alices select randomly some bits and announce their selection.
  All Bobs and all Alices compare  the values of these check  bits. If the test passes, then the rest of the unchecked bits of
   $\{\oplus^n_{l=1}(a^1_{jl}\oplus a^2_j\oplus\cdots\oplus a^m_j)\}^N_{j=1}$
 are the raw key for secret sharing between two groups. We should emphasize that $n$ must be odd in case (B)
since $\oplus^n_{l=1}(a^1_{jl}\oplus a^2_j\oplus\cdots\oplus
a^m_j)=a^1_j\oplus na^2_j\oplus\cdots\oplus na^m_j=a^1_j$ if $n$
is even.

\section {security}

Now we  discuss the unconditional security of this quantum secret sharing protocol between $m$ parties and $n$
parties. Note that the encoding of secret messages  by  Alice$i$ ($1\leq i\leq m$) is identical to the process
in a one-time-pad encryption where the text is encrypted with a random key as the state of the  photon in the
protocol is completely random. The great feature of a one-time-pad encryption is that as long as the key strings
are truly secret, it is  completely safe and no secret messages can be leaked even if the cipher-text is
intercepted by the eavesdropper. Here the secret sharing protocol is even more secure than the classical
one-time-pad in the sense that an eavesdropper Eve can not intercept the whole cipher-text as the photons'
measuring-basis is chosen randomly. Thus the security of this secret sharing protocol depends entirely on the
second part when Alice$m$ sends the $l$-th sequence of $N$ photons  to Bob$l$ ($1\leq l\leq n$).

 The process for ensuring a secure block of $nN$ qubits ($n$ secure sequences of  $N$ photons) is similar to
 that in the BB84 QKD protocol
 \cite {BB84}. The process of this secret sharing between $m$ parties and $n$ parties after all Alices encoding
 their respective messages using unitary operations is in fact identical to $n$ independent BB84 QKD processes,
 which has been proven unconditional secure \cite {SP, CBKG}.  Thus the security for the present quantum  secret sharing
 between multi-party and multi-party is guaranteed.

In practice,  some qubits may be lost in transmitting. In this case, all Alices and Bobs can take two kind
strategies, one is  removing these qubits, the other is using a qubit chosen at random in one of four states
$\{|0\rangle, |1\rangle, |+\rangle, |-\rangle\}$ as a substitute for a lost qubit.  If a member  does not
receive a qubit and wants to delete it, she/he must announce  and let all members in the two groups know the
fact.  All Alices and all Bobs  sacrifice some randomly selected qubits to test the "error rate". If the error
rate is too high, they abort the protocol. Otherwise, by utilizing a Calderbank-Shor-Steane (CSS) code \cite{CS,
Steane, SP}, they perform information reconciliation and privacy amplification on the remaining bits to obtain
secure final key bits for secret sharing. They proceed to this step  obtaining the final key  while all Alices
communicate with all Bobs. In a CSS mode, classical linear codes $C_1$ and $C_2^\perp$ are used for bit and
phase error correction, respectively, where $C_2\subset C_1$. The best codes that we know exist satisfy the
quantum Gilbert-Varshamov bound.  The number of cosets of $C_2$ in  $C_1$ is $|C_1|/|C_2|=2^M$ so there is
 a one-to-one correspondence $u_K\rightarrow K$ of the set of representatives $u_K$ of  the $2^M$ cosets
of $C_2$ in $C_1$ and the set of M-bit strings  $K$. As in the BB84 protocol, $C_1$ is used to correct bit
errors in the key, and $C_2$ to amplify privacy. For the sake of convenience, we suppose that after verification
test all Alices are left with the $N'$ bit string $v=\{\oplus^n_{l=1}(\oplus^m_{i=1}a_{n+l}^i)$,
$\oplus^n_{l=1}(\oplus^m_{i=1}a_{2n+l}^i), \cdots$,$
\oplus^n_{l=1}(\oplus^m_{i=1}a_{nN'+l}^i)\}=\{\oplus^n_{l=1}(\oplus^m_{i=1}a_{ns+l}^i)\}_{s=1}^{N'}$, but all
Bobs with $v+\epsilon$ by the effect of losses and noise.    Let us assume that {\it a priori} it is known that
along the communication channel used by all Alices and all Bobs, the expected number of errors per block caused
by losses and all noise sources including eavesdropping is less than $t=\delta N'$, where $\delta$ is the bit
error rate. How can  an upper bound be placed on $t$? In practice, this can be established by random testing of
the channel, leaving us with a protocol which is secure \cite{GLLP}, even against collective attacks. If
$\delta$ is low enough, we can be confident that error correction will succeed, so that all Alices and all Bobs
share a secure common key. The secure final key for secret sharing can be extracted from the raw key bits
(consisting of the remaining noncheck bits) at the asymptotic rate $R=\mathrm{Max}\{1-2H(\delta), 0\}$
\cite{GLLP}, where $\delta$ is the bit error rate found in the verification test (assuming $\delta<1/2$). Using
a pre-determined $t$ error-correcting CSS code \cite{SP}, the two groups share a secret key string and realize
secure communication. Suppose that government department1 wishes to send messages to government department2,
then all Alices gather together, choose a random code word $u$ in $C_1$ ($u$ may be $u_1+u_2+\cdots +u_m$, where
$u_i$ is a code word in $C_1$ selected randomly by Alice$i$), and encode their $M$-bit message $P$ by adding the
message and the $M$-bit string $K$ together, where $u+C_2=u_K+C_2$, then they send it to government department2.
Bobs receive the secret message and publicly announce this fact. All Alices announce $u+v$. All Bobs subtract
this from their result $v+\epsilon$, and correct the result $u+\epsilon$ with code $C_1$ to obtain the code word
$u$. All Alices and all Bobs use the $M$-bit string $K$ as the final key for secret sharing. That is, all Alices
and all Bobs perform information reconciliation by the use of the classical code $C_1$, and performs privacy
amplification by computing the coset of $u+C_2$. All Bobs can decode and read out the message $P$  by
subtracting $K$. No one in department1 tells final key $K$ to someone or part of  department2, since the aim of
all Alice is to let all Bobs know their message.

 In summary, we propose a  scheme for quantum secret sharing between multi-party and multi-party, where  no
 entanglement  is employed.
In the protocol,  Alice1 prepares a sequence of single photons in
one of four different states according to her two random bits
strings, other Alice$i$ ($2\leq i\leq m$) directly encodes her two
random classical information strings on the resulting sequence of
Alice$(i-1)$ via unitary operations, after that Alice$m$ sends
$1/n$ of the sequence of single photons to each Bob$l$ ($1\leq
l\leq n$). Each Bob$l$ measures his photons according to all
Alices' measuring-basis sequences. All Bobs must cooperate in
order to infer the secret key shared by all Alices. Any subset of
all Alices or all Bobs can not extract secret information, but the
entire set of all Alices and the entire set of all Bobs can.
      As entanglement, especially the inaccessible multi-party entangled state, is not
 necessary in the present quantum secret sharing protocol between $m$-party and $n$-party, it may be more
 applicable when the numbers $m$ and $n$ of the parties of secret sharing are large. Its theoretic efficiency is
 also doubled to approach $100\%$. This protocol is feasible with present-day techniques.

 \acknowledgments This work was supported by the Hebei Natural Science Foundation of China under Grant Nos:
A2004000141 and A2005000140 and the Key Natural Science Foundation of Hebei Normal University.

\end{document}